\documentclass{article}

\usepackage{graphicx}
\usepackage{amsmath, amssymb}
\usepackage[mathscr]{eucal}

\newcommand{\bra}[1]{\langle\,#1\,|}
\newcommand{\ket}[1]{|\,#1\,\rangle}
\newcommand{\p}{\partial}
\newcommand{\ud}{\mathrm{d}}
\newcommand{\pathD}{\!\mathscr{D}}
\newcommand{\Det}{\text{Det}\,}
\newcommand{\x}{{\bf x}}
\newcommand{\y}{{\bf y}}

\newcommand{\q}{{\bf q}}

\newcommand{\X}{\mathbf{X}}

\newcommand{\BB}{\mathbf{B}}

\newcommand{\til}{\widetilde}

\title{Time evolution in string field theory and T-Duality}

\author{Anton Ilderton\footnote{a.b.ilderton@dur.ac.uk, corresponding author}\hspace{2pt} and Paul Mansfield\footnote{p.r.w.mansfield@dur.ac.uk} \\  \\ Centre for Particle Theory, University of Durham, \\Durham, DH1 3LE, UK}	
\date{}

\begin{document}
\maketitle
\section*{Introduction}

In the mechanics of particles and fields it is natural to consider the evolution in time of arbitrary configurations. In second quantised string theory this is not so straightforward, for example in Witten's theory \cite{Witten} the natural time variable is that at the mid-point of the string rather than a global time for the whole string. In this letter we will construct the time evolution operator for second quantised strings by analogy with that for field theory. We begin by showing that when the field theory Schr\"odinger functional is written in terms of propagators expressed in first quantised form then these describe particles  moving on a timelike orbifold $\mathbb{S}^1/\mathbb{Z}_2$. The first quantised propagators have an immediate generalisation to string theory, suggesting that the Schr\"odinger functional for second quantised strings can be expressed in terms of first quantised strings moving on this orbifold. To strengthen the analogy we give a graphical construction of the field theory Schr\"odinger functional which extends to both open and closed string theory.  This avoids using a Lagrangian formulation of string field theory. Finally we study the effect of T-duality on time evolution and describe the nature of BRST invariance in our approach.

\section{Time Evolution in QFT}

Consider a bosonic scalar field $\phi$ in $D+1$ dimensions.
It will be convenient to work in a basis in which $\hat\pi(\x)=\dot\phi(\x)$,
the momentum canonically conjugate to the field (the problem of defining a momentum in string field theory goes hand in hand with the definition of a global time), is diagonal
\begin{equation*}
  \bra{\pi}\hat\pi(\x) = \pi(\x)\bra{\pi}, \quad i\frac{\delta}{\delta \pi(\x)}\bra{\pi} = \bra{\pi}\hat{\phi}(\x).
\end{equation*}
Symanzik has shown how to express the Schr\"odinger functional
in the representation in which the field is diagonal as a 
functional integral \cite{Symanzik} using sources. Generalising this to the momentum
representation gives
the Schr\"odinger functional as
\begin{equation*}
    \mathscr{S}[\pi_2,t_2;\pi_1,t_1] = \bra{\pi_2}e^{-i\hat{H}(t_2-t_1)}\ket{\pi_1} = \int\pathD\varphi\,e^{iS[\varphi] +i\int\pi_2\varphi(t_2)-i\int\pi_1\varphi(t_1)}\bigg|_{\dot\varphi(t_1)=0}^{\dot\varphi(t_2)=0}.
\end{equation*}
This has a Feynman diagram expansion in propagators which obey Neumann boundary conditions on the boundaries at times $t_1$ and $t_2$, where all external legs must end, and vertices integrated over the interval. The free field contribution is
\begin{equation}\label{schro-diags}
    \includegraphics[width=0.9\textwidth]{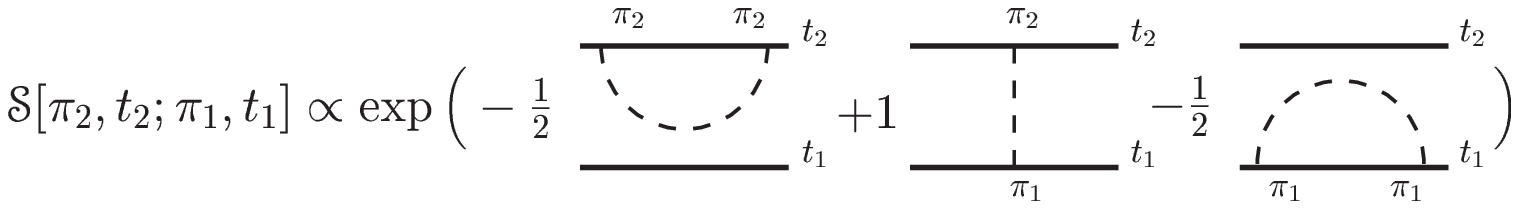}
\end{equation}
where the broken line represents the propagator, which we call $G_\text{orb}$, and the heavy lines are the spacelike boundaries. We discuss the normalisation coming from the Gaussian integral below. Without loss of generality, we will take $t_1=0$ and $t_2=t$ from here on. The required boundary conditions on the propagator can be achieved using the method of images,
\begin{equation}\label{orb-def}
  G_\text{orb}(\x,t_f;\y,t_i) = \sum\limits_{n\in\mathbb{Z}}G_0(\x,t_f+2nt;\y,t_i) +\sum\limits_{n\in\mathbb{Z}}G_0(\x,-t_f+2nt;\y,t_i),
\end{equation}
for $0\leq t_i,t_f\leq t$ and $G_0$ is the free space propagator.  To interpret this in terms of first quantisation recall that $G_0$ is given by a sum over paths $x(\xi)$ with an action involving an intrinsic metric $g$,
\cite{Brink}. Integrating out $g$ gives
a Boltzmann weight equal to the exponential of the length of the path,
\begin{equation}\label{prop-path}\begin{split}
  G_0(x_f; x_i) &= \int \pathD (x,\,g)\,\, e^{i\int_0^1 d\xi\,\, ({\dot x\cdot\dot x}/(2g)+m^2g/2)}\bigg|_{x(0)=x_i}^{x(1)=x_f} \\
  & =\int \pathD x\,\, e^{im\int_0^1 d\xi\,\, \sqrt{\dot x\cdot\dot x}}\bigg|_{x(0)=x_i}^{x(1)=x_f}.
\end{split}\end{equation}
To obtain $G_\text{orb}$, we identify free space points with their images under an $\mathbb{S}^1/\mathbb{Z}_2$ (orbifold) compactification of the time direction, with radius $t/\pi$. The sum over paths to each image gives a free propagator in the sum (\ref{orb-def}).

The first form of the functional integral in (\ref{prop-path}) is immediately generalised to string theory suggesting that the Schr\"odinger functional for second quantised string theory can be obtained by letting the propagators in (\ref{schro-diags}) represent the string propagator on the orbifold. It is not obvious how to derive this from a Lagrangian  given the remarks in the introduction about the r\^ole of a global time in Witten's open string field theory, and given the difficulties of closed string field theory (for a review see \cite{Zwei} and references therein). Rather than attempt a Lagrangian derivation we will give a graphical derivation of the field theory result which can be
taken over into string theory.

We appeal to a fundamental property of field theory, which follows from the observation that paths from $t_3$ to $t_1$ must cross the plane at time $t_2$ for $t_3>t_2>t_1$, so that formally the sum over paths in (\ref{prop-path}) can be factorised,
\begin{equation*}
\sum_{\textrm{paths AB}} e^{-\textrm{length(AB)}}= \sum\limits_\textrm{C}  \bigg( \sum_{\textrm{paths AC}}
e^{-\textrm{length(AC)}}\bigg) \bigg( \sum_{\textrm{paths CB}} e^{-\textrm{length(CB)}}\bigg). 
\end{equation*}
The explicit result, which we refer to as the gluing property, is
\begin{equation}\label{factorisation}
  \int\!\ud^D \y \,\, G_0(\x_3,t_3;\y_2,t_2)\bigg(-i\overleftrightarrow{\frac{\partial}{\partial t_2}}\bigg)G_0(\y,t_2;\x_1,t_1)= G_0(\x_3,t_3;\x_1,t_1),
\end{equation}
for $t_3>t_2>t_1$. More generally, if the endpoints $x_3$ and $x_1$ are on opposite sides of the plane at time $t_2$, the propagators are glued to form the usual propagator. If they are on the same side gluing produces the image propagator $G_I$ equal to the free space propagator for the points $x_3$ and the reflection of $x_1$ in the plane at $t_2$. The cases are summarised below,
\begin{equation}\label{rules}
\int\!\ud^D \y \,\, G_0(\x_2,t_2;\y,t)\frac{\p}{\p t}G_0(\y,t;\x_1,t_1)=
\begin{cases}
  \mp \frac{i}{2}G_0(\x_2,t_2;\x_1,t_1)& t_2\gtrless t\gtrless t_1 \\ 
  \mp \frac{i}{2}G_I(\x_2,t_2;\x_1,t_1)& t\gtrless t_1, t_2 \\
\end{cases}
\end{equation}
Applying  (\ref{rules}) twice we obtain
\begin{equation*}
\begin{split}
\int\ud^D(\x_3,\x_2)\,\, &G_0(\x_4,t_4,\x_3,t_3)
\bigg( 4\frac{\p^2}{\p t_3\, \p t_2 }\,G_0(\x_3,t_3,\x_2,t_2)\bigg)
G_0(\x_2,t_2,\x_1,t_1)\\
&= G_0(\x_4,t_4,\x_1,t_1)\quad\text{for}\quad t_4>t_3>t_2>t_1.
\end{split}
\end{equation*}
Taking all the $t_i$ to zero gives a useful relation which may be expressed as 
\begin{equation*}
  \includegraphics[width=0.5\textwidth]{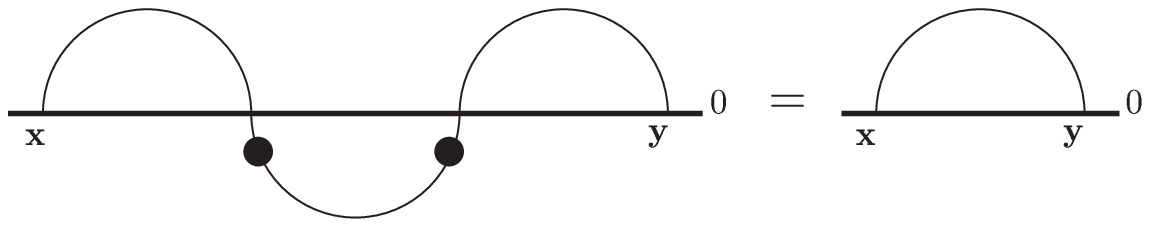}
\end{equation*}
where the heavy line is the plane at time $t=0$, the unbroken line is the free space propagator and a black dot is $-2$ times a time derivative. Thus
\begin{equation}
  \includegraphics[width=0.4\textwidth]{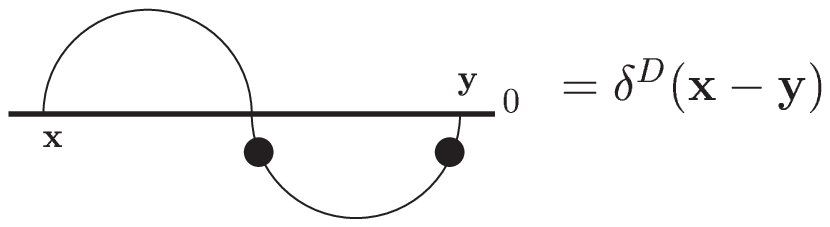}
\end{equation}
From this we deduce that the inverse of the free space propagator at equal time is
\begin{equation}\label{inverse}
  \includegraphics[width=0.15\textwidth]{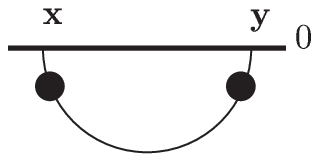}.
\end{equation}
We can now show that time evolution is captured by the gluing rules and the Feynman diagram expansion. Consider calculating the free theory two-point function at unequal times,
\begin{equation}\label{2point}
  \langle\,\pi(\x,t)\pi(\y,0)\,\rangle = \int\pathD(\pi_2,\pi_1) \Psi_0[\pi_2]\pi_2(\x)\mathscr{S}[\pi_2,t;\pi_1,0]\pi_1(\y)\Psi_0[\pi_1].
\end{equation}
The vacuum wave functional $\Psi_0[\pi]$ can be constructed by requiring that it yield $G_0$ at equal times as a vacuum expectation value,
\begin{equation*}
  G_0(\x,0;\y,0) = \langle\,\phi(\x,0)\phi(\y,0)\,\rangle = -\int\pathD\pi \, \Psi_0[\pi]\frac{\delta}{\delta\pi(\x)}\frac{\delta}{\delta\pi(\y)}\Psi_0[\pi] \\
\end{equation*}
\begin{equation}
\includegraphics[width=0.4\textwidth]{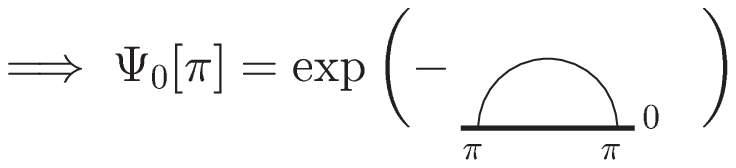}.
\end{equation}
The $\pi_1$ integration in (\ref{2point}) is Gaussian in the free theory,
\begin{equation*}
  \includegraphics[width=0.8\textwidth]{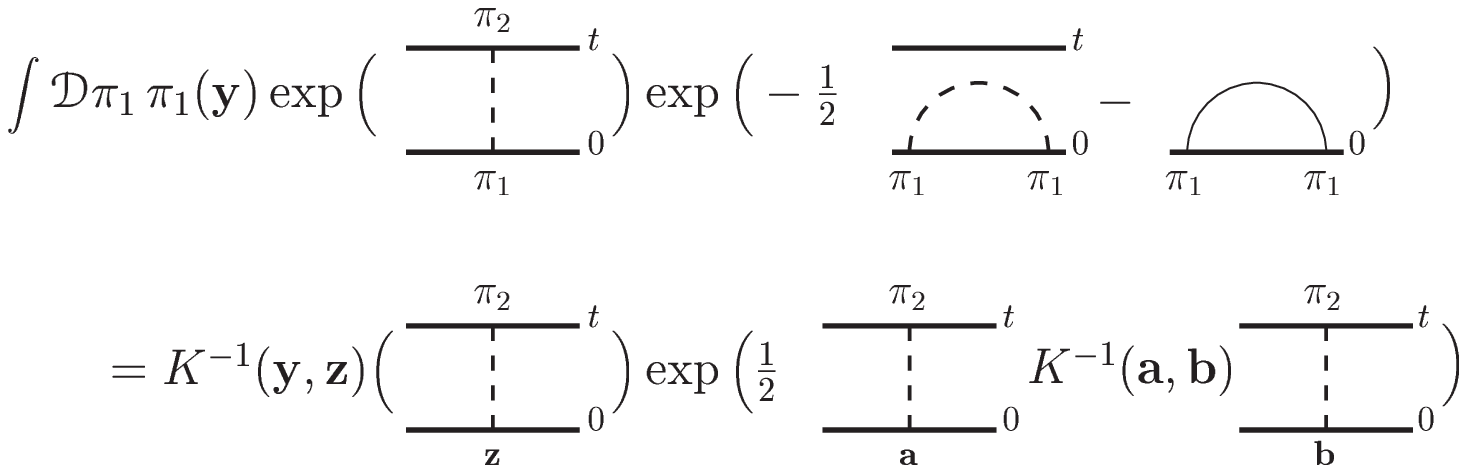}.
\end{equation*}
The symmetric operator $K$ and it's inverse are
\begin{equation*}
  \includegraphics[width=0.85\textwidth]{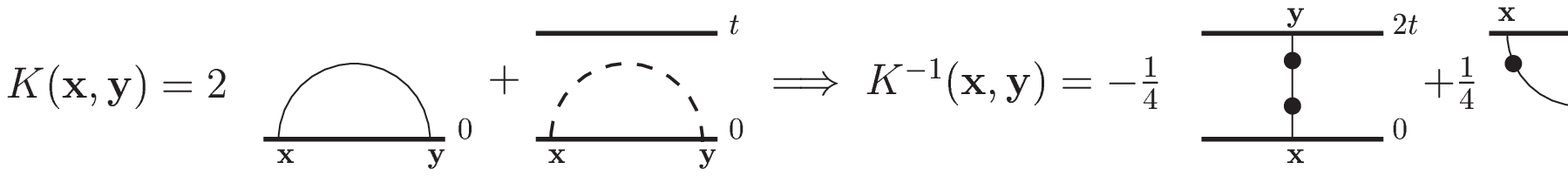}
\end{equation*}
which can be checked using the gluing rules (\ref{rules}) and the corollary (\ref{inverse}). The exponential term is
\begin{equation*}
  \includegraphics[width=0.9\textwidth]{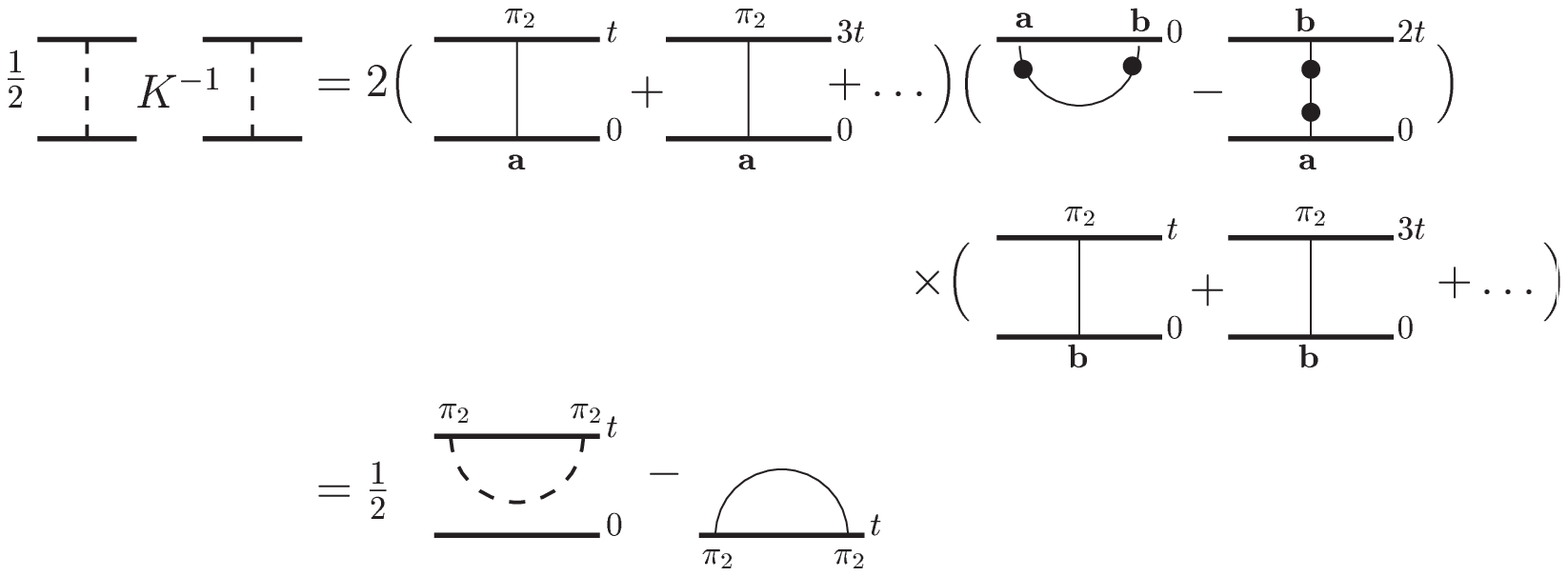}
\end{equation*}
which leaves us with another Gaussian integral in $\pi_2$,
\begin{equation*}
  \includegraphics[width=0.9\textwidth]{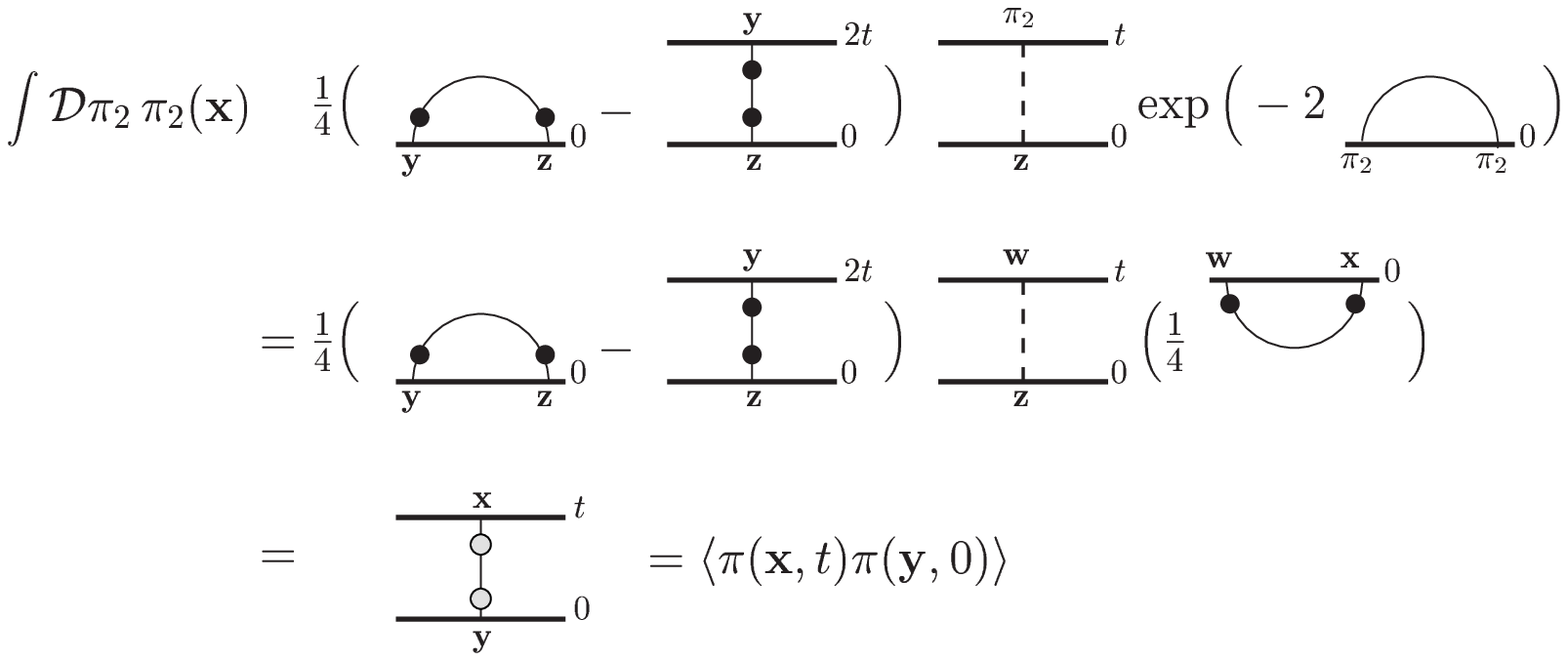}
\end{equation*}
We arrive at the Feynman diagram of the free propagator with the correct coefficient of unity; the grey dot denotes a time derivative (with no factor) which appears since we are computing correlation functions of $\pi=\dot\phi$.

So using the gluing property alone we have shown that the expression (\ref{schro-diags}) for the Schr\"odinger functional leads to the correct result for the two point function at unequal times. This argument is invertible; If we know the two point function we can construct the Schr\"odinger functional as in (\ref{schro-diags}) provided the gluing property holds. If we can generalise the gluing property to string theory we can repeat the diagrammatic arguments and construct the second quantised string Schr\"odinger functional.

\section{The String Field Propagator}

The string field propagator can be constructed in much the same way as in QFT \cite{Brink}, \cite{Polyakov} as the transition amplitude $G(X_f;X_i)$ between arbitrary spacetime curves $X_i(\sigma)$ and $X_f(\sigma)$. We denote the propagator with boundary conditions $X^0(\sigma)=$ constant, between arbitrary spacelike curves $\X(\sigma)$ as
\begin{equation}\label{prop-trans}
  G_{t_f-t_i}(\X_f; \X_i) = \int\pathD(X,g)\,\, e^{-\frac{1}{4\pi\alpha'}\int\!\ud^2\sigma\sqrt{g}\,  g^{ab} \partial_a X^\mu \partial_b X_\mu}\bigg|_{\X=\X_i(\sigma),\,\, X^0=t_i.}^{\X=\X_f(\sigma),\,\, X^0=t_f}
\end{equation}
At tree level the worldsheet is a finite strip (cylinder) for open (closed) strings. An arbitrary metric can be written as a diff$\times$Weyl transformation (orthogonal to the CKV for the closed string) of a reference metric $\hat g_{ab}(T)$ for some value of the Teichm\"uller parameter $T$. We take $\hat g_{ab}(T)=\text{diag}(1,T^2)$ so that $T$ represents the intrinsic length of the worldsheet. The propagator is \cite{Moore}
\begin{equation}\label{prop-def}
  G_t(\X_f; \X_i) = \int\limits_0^\infty\!\ud T\, \text{Jac}(T)(\Det' \widehat{P}^\dagger \widehat{P})^{\frac{1}{2}}(\Det\widehat{\Delta})^{-13}\int\pathD\xi\, e^{-S_\text{cl}[X_\text{cl},\hat{g}(T)]}.
\end{equation}
The measure on Teichm\"uller space is Jac$(T)$ given by
\begin{equation*}
\text{Jac}(T)_\text{open}=\frac{(h_{ab}|\chi_{ab})}{(h_{ab}|h_{ab})^{1/2}}, \quad \text{Jac}(T)_\text{closed}=\frac{(h_{ab}|\chi_{ab})}{(V^a|V^a)^{1/2}(h_{ab}|h_{ab})^{1/2}}
\end{equation*}
where $h_{ab}$ is the zero mode of $\hat{P}^\dagger$, $\chi_{ab}$ is the symmetric traceless part of $\hat{g}_{ab,T}$ and $V^a$ is the CKV on the cylinder. $X_\text{cl}$ satisfies the wave equation in metric $\hat{g}$ with boundary conditions ${X^{\xi}_\text{cl}}|_{\tau=0} = X_i(\sigma)$, ${X^{\xi}_\text{cl}}|_{\tau=1} = X_f(\sigma)$. The remaining $\xi$ integral is over reparametrisations of the boundary data. If we attach reparametrisation invariant functionals $\Pi_i[\X_i]$, $\Pi_f[\X_f]$ to the boundaries of the worldsheet then this integral can be done trivially to give an (infinite) constant factor, for then
\begin{equation}
\begin{split}
  \int\pathD(X_f, X_i)&\int\pathD\xi\, e^{-S_\text{cl}[X_\text{cl},\hat{g}]}\,\,\Pi_i[X_f]\Pi_f[X_i] \\
  &= \int\pathD\big({X_\text{cl}}|_{\tau=1},{X_\text{cl}}|_{\tau=0}\big)\,\,e^{-S_\text{cl}[X_\text{cl},\hat{g}]}\,\,\Pi_f[{X_\text{cl}}|_{\tau=1}]\Pi_i[{X_\text{cl}}|_{\tau=0}]\int\pathD\xi.
\end{split}
\end{equation}
The same applies when we sew two worldsheets together, since $G$ itself is reparametrisation invariant. Carlip's sewing method \cite{Carlip}, required for correctly combining moduli spaces, involves integrating over all boundary values of $X^0$ which in our problem is not appropriate. We wish to generalise (\ref{factorisation}), the key to which is the correct identification of the degrees of freedom on the boundary. The Alvarez conditions \cite{Alvarez} on the reparametrisations,
$n^a\xi_a = n^a t^bP(\xi)_{ab} =0$, split into orthogonal pieces on the strip (cylinder), the $\tau$- components of which do not couple to reparametrisations of the boundary. When we sew two worldsheets only half of the determinant of $P^\dagger P$ is sewn together, the remainder cancelling the effects of not integrating over $X^0$.

We can make this precise using ghosts. The r\^ole of the ghosts in string theory is to cancel the undesirable effects of including the $X^0$ oscillators. Our ghosts will do the same thing, although to a different end. Take the usual representation of the metric integral
\begin{equation}
  \big(\Det' P^\dagger P\big)^{1/2} = \int\pathD(b,c) e^{-\frac{1}{2}\int b_{ab}P(c)^{ab}}
\end{equation}
and change variable $b=P\gamma$ for Grassmann vector $\gamma$, which turns the ghost sector of the path integral into
\begin{equation}
  \big(\Det' P^\dagger P\big)^{1/2} = \int\pathD(\gamma,c) \big(\Det' P^\dagger P\big)^{-1/2}e^{-\frac{1}{2}\int P(\gamma)_{ab}P(c)^{ab}}.
\end{equation}
Now represent the determinant on the RHS of the above by a bosonic vector integral,
\begin{equation}
  \big(\Det' P^\dagger P\big)^{-1/2} = \int\pathD f\, e^{-\frac{1}{2}\int P(f)_{ab}P(f)^{ab}}.
\end{equation}
For the closed string this change of variable is defined only up to shifts $J^a\rightarrow J^a+\lambda V^a$ for $J^a\in\{c^a,\gamma^a,f^a\}$, so we choose $(J|V)=0$ removing the c.o.m. from the classical pieces. This is our new ghost system. In accord with the Alvarez conditions we integrate out the fields $J^\tau$ and fix the values of $J^\sigma$ on the boundaries so the propagator in the extended Hilbert space interpolates between arbitrary values of $\X(\sigma), \{J^\sigma(\sigma)\}$ at particular times. Letting $\BB$ denote boundary values of $\mathbf{X}^i$, the set $J^\sigma$ and the $X^0$ oscillators (zero) it can be shown \cite{us} that the Euclidean generalisation of (\ref{factorisation}) holds in string theory as
\begin{equation}\label{string-sew}
  \int\!\pathD \BB \,\, G_{t_2-t}(\BB_2;\BB)\,\overleftrightarrow{\frac{\partial}{\partial t}}\,G_{t-t_1}(\BB;\BB_1)= G_{t_2-t_1}(\BB_2;\BB_1),\quad t_2> t>t_1.
\end{equation}

We can check our method using the cancellation of the Weyl anomaly. Even in the critical dimension (\ref{prop-trans}) has a dependence on the Liouville field at the corners of the open string worldsheet \cite{Paul-ghosts}. The sewing prescription we have described cancels the anomaly on the boundaries being sewn, ensuring that the sewn worldsheet carries no anomaly in the bulk.

Our factorisation of the ghosts may seem ad-hoc but in fact follows from the gauge choice
\begin{equation}\label{GF-conds}
  \int\!\ud^2\sigma\sqrt{g}\,g^{ab}\hat{h}_{ab}\equiv(\hat{h}_{ab}|g_{ab}) =0,\qquad \hat{P}^\dagger\bigg(\frac{\sqrt{g}g^{rs}}{\sqrt{\hat{g}}}\bigg)^a =0,
\end{equation}
where a hat denotes use of the fiducial metric (this is equivalent to the usual choice $\sqrt{g}g^{ab} = \sqrt{\hat{g}}\hat{g}^{ab}(T)$). The corresponding gauge fixed action is
\begin{equation}\label{theact}
  S_\text{BRST} = \frac{1}{2}\int\!\ud^2\sigma\sqrt{\hat{g}}\hat{g}^{ab}\p_a X\p_b X + \frac{1}{2}\int\!\ud^2\sigma\sqrt{\hat{g}}\,\,\hat{P}(\gamma)_{ab}\hat{P}(c)^{ab}.
\end{equation}
The BRST transformations are 
\begin{equation}\label{BRST-trans2}
  \delta_Q X = c^a\partial_a\, X, \quad \delta_Q c^a = c^b\partial_b\, c^a, \quad \delta_Q \hat{P}(\gamma)_{ab} = -2\hat{T}_{ab},
\end{equation}
where $\hat{T}_{ab}$ is the usual string energy momentum tensor with $b_{ab} = \hat{P}(\gamma)_{ab}$. This set of transformations are non-local but have the natural interpretation of generating local reparametrisations of the boundary. Consider the string field propagator written as
\begin{equation}
  G(\X_f;\X_i) = \int\pathD(X,\gamma,c)(\Det\hat{P}^\dagger \hat{P})^{-1/2}e^{-S_\text{BRST} -S_J}\bigg|_{X=X_i}^{X=X_f}
\end{equation}
where $S_\text{BRST}$ is as in (\ref{theact}) and $S_J$ is a source term which generates boundary values of the ghosts,
\begin{equation}\label{source}
  S_J = \frac{1}{2}\int\!\ud^2\sigma\sqrt{\hat{g}}\,\,  (\hat{P}^\dagger\hat{P}\gamma)_ac^a_\text{cl} + \gamma^a_\text{cl}(\hat{P}^\dagger\hat{P}c)_a
   = \int\limits_\text{bhd}\!\ud\Sigma^s\,\,  (\hat{P}\gamma)_{rs}c^s_\text{b} + \gamma^s_\text{b}(\hat{P}c)_{rs}.
\end{equation}
In the above, $c^a_\text{cl}$ obeys $\hat{P}^\dagger\hat{P}=0$ and equals the boundary values of the ghosts on the Dirichlet sections of the worldsheet. The bulk action $S_\text{BRST}$ is BRST invariant for arbitrary boundary values of $X,c,\gamma$. The source term $S_J$ does not respect this symmetry, so we are led to expect a Ward identity resulting from a shift in integration variables corresponding to (\ref{BRST-trans2}), $\langle\,\delta_Q S_J\,\rangle = 0$. In addition to the usual short distance divergences in the quantum pieces, the corner anomaly leads to finite, non-zero contributions from the image charges. We find the Ward identity can be written as an operator acting on the propagator,
\begin{equation}
\bigg[\int\limits_0^\pi\!\ud\sigma\,\, c_\text{b}^\sigma \X_\text{b}'\frac{\delta}{\delta \X_\text{b}} + \big(c^\sigma_\text{b}\gamma^\sigma_\text{b}\big)'\frac{\delta}{\delta \gamma^\sigma_\text{b}} +\frac{1}{2} c^\sigma_\text{b} {{c^\sigma}'_\text{b}}\frac{\delta}{\delta c^\sigma_\text{b}} +\frac{26}{8}\big({c^\sigma}'(0)+{c^\sigma}'(\pi)\big)\bigg]G=0,
\end{equation}
where a subscript $\text{b}$ indicates boundary data. This operator describes the transformation of 25 scalars $\X$ and the tangential component of a vector $\gamma^\sigma$ under a reparametrisation of the boundary generated by the ghost $c^\sigma$, with quantum corrections. Demanding BRST invariance here does not put the string field on shell, as the reparametrisations are only a subset of those described by BRST in the usual formalism.

\section{T-Duality in the Schr\"odinger Functional}
The sewing rule (\ref{string-sew}) confirms that despite the extended nature of the string a Schr\"odinger representation makes sense for string field theory, and we can carry over our diagrammatic arguments so that
\begin{equation}\label{string-schro}
  \includegraphics[width=0.8\textwidth]{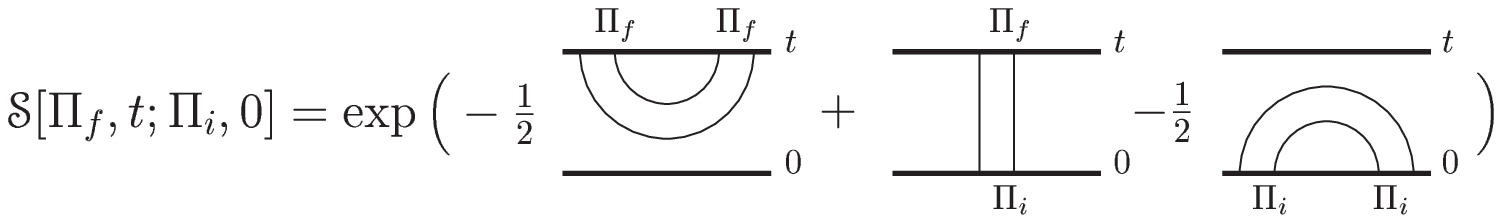}
\end{equation}
for momentum string fields $\Pi[\X, {J^\sigma}]$, and the double line represents the orbifolded propagator for either the open or closed string. The normalisation constant is discussed below. The orbifold leads naturally to the question of what r\^ole T-duality plays. We now show that T-duality exchanges the states attached to the propagators with backgrounds in the dual picture, and vice versa. 

The closed string Schr\"odinger functional is T-dual to the loop diagrams appearing in the normalisation of the open string Schr\"odinger functional. We set $t=\pi R$, making the orbifold radius explicit, and use Poisson resummation and a change in modular parameter to convert the closed propagators into open loops so that, in an obvious notation, the Schr\"odinger functional becomes
\begin{equation}\label{logs}\begin{split}
  \log \mathscr{S}_\text{closed} &= -\frac{1}{2}\sum_\text{$n$ even}\!\Pi_f G_{\pi Rn}\Pi_f +\sum_\text{$n$ odd}\!\Pi_f G_{\pi Rn}\Pi_i-\frac{1}{2}\sum_\text{$n$ even}\!\Pi_i G_{\pi Rn}\Pi_i \\
  &= -\frac{1}{2}\sum_\text{$n$ even}\Pi_f G_{\pi\til{R}n}\Pi_f +\!\!\sum_\text{$n$ even}\Pi_i G_{\pi\til{R}n}e^{in\pi/2}\Pi_f-\frac{1}{2}\sum_\text{$n$ even}\Pi_i G_{\pi\til{R}n}\Pi_i
\end{split}\end{equation}
where $\til{R}=\alpha'/R$ and $G$ in the second line is an open string contribution. The new exponent comes from the resummation and the fields glued onto the Dirichlet sections of the closed string propagator become an averaging over backgrounds characterised by $\Pi_i$, $\Pi_f$ coupling to the ends of the open string. These backgrounds, as they must be, are the same at each end of the string, for we can write the above as
\begin{equation}
  \log\mathscr{S}_\text{closed} = -\frac{1}{2}\sum\limits_\text{$n$ even}\big(\Pi_i-e^{iA\int\!\ud X^0}\Pi_f\big) G_{\pi\til{R}n} \big(\Pi_i-e^{iA\int\!\ud X^0}\Pi_f\big)
\end{equation}
with Wilson line value $A=(2\til{R})^{-1}$. Let us give an explicit example. We take reparametrisation invariant boundary states $\Pi_{\,i,f}[\X]=\delta^p(\X(\sigma)-\q_{\,i,f})$ for $\q_{\,i,f}$ constant $p$-vectors. These are pointlike states in $p$ directions and Neumann states in $25-p$ directions, $0\leq p\leq 25$. The closed string Schr\"odinger functional is
\begin{equation}\label{example1}\begin{split}
\log \mathscr{S}_\text{closed}= &-\text{Vol}^{25-p}\int\limits_0^\infty\!\frac{\ud T}{T^\frac{p+1}{2}}\,\,e^{2T}\prod\limits_{m=1}\big(1-e^{-2mT}\big)^{-24}\sum\limits_\text{$n$ even} e^{-\frac{\pi^2 R^2}{2\alpha'T}n^2} \\
&+\text{Vol}^{25-p}\int\limits_0^\infty\!\frac{\ud T}{T^\frac{p+1}{2}}\,\,e^{-\frac{\delta\q^2}{2\alpha' T} + 2T}\prod\limits_{m=1}\big(1-e^{-2mT}\big)^{-24}\sum\limits_\text{$n$ odd} e^{-\frac{\pi^2 R^2}{2\alpha'T}n^2}
\end{split}\end{equation}
with $\delta\q=\q_f-\q_i$. Since the open string runs from $\sigma=0\ldots\pi$ and the closed string from $\sigma=0\ldots 2\pi$ we must scale the closed string worldsheet to interpret (\ref{example1}) as an open loop. We include this in a change of modular parameter $S:=2\pi^2/T$. After this and a Poisson resummation we find
\begin{equation}\label{example2}\begin{split}
  \log\mathscr{S}_\text{closed}=-\text{Vol}^{25-p}\int\limits_0^\infty\frac{\ud S}{S}\frac{1}{S^\frac{26-p}{2}}e^S&\prod\limits_{m=1}(1-e^{-mS})^{-24} \\
  &\times\sum\limits_\text{$n$ even}e^{-\frac{\pi^2\til{R}^2}{4\alpha'S}n^2}\bigg(1-e^{\frac{in\pi}{2}} e^{-\frac{\delta\q^2 S}{4\pi\alpha'}}\bigg).
\end{split}\end{equation}
Now consider an open string loop. The measure on Teichm\"uller space is $\ud S/S$ (this gives the logarithm of the worldsheet propagator). If the string has Neumann conditions in $26-p$ directions (including $X^0$) and Dirichlet conditions in $p$ directions, as for a string on a D$(25-p)$-brane then the trace over $\X$ gives the eta function and the factor $(S^{-1/2}\text{Vol})^{(25-p)}$ from the $25-p$ zero modes. The sum and remaining factor of $S^{-1/2}$ come the trace over $X^0$ in the co-ordinate representation. We arrive at (\ref{example2}), if the term in large brackets represents an averaging over backgrounds of Wilson lines and D$(25-p)$-branes of separation $\delta\q$.

We interpret the open string duality as taking us from one Schr\"odinger functional to another with an exchange of boundary states and backgrounds. Explicit examples are difficult to construct since the corner anomaly, not an issue for the closed string,  forces us to find conformally invariant states and backgrounds, but we can give an outline. Poisson resummation implies
\begin{equation}
  \bigg(\frac{\pi R^2}{\alpha' T}\bigg)^{1/2}\sum\limits_\text{$n$ even}e^{-\frac{\pi^2 R^2}{4\alpha' T}n^2} = \sum\limits_\text{$n$ even}e^{-\frac{\overline{R}^2 T}{4\alpha'}n^2} + \sum\limits_\text{$n$ odd}e^{-\frac{\overline{R}^2 T}{4\alpha'}n^2}
\end{equation}
(for the odd sum the plus on the R.H.S. becomes minus) where $\overline{R}=2\alpha'/R$. Following a modular transformation $S:=\pi^2/T$ these are the sums in the open string Schr\"odinger functional. Again the states now represent an averaging over backgrounds. The open string Schr\"odinger functional becomes
\begin{equation}
  \log\mathscr{S}_\text{open} = -\frac{1}{2}\sum_\text{$n$ even}(\Pi_i - \Pi_f) G_{n\pi\overline{R}}(\Pi_i-\Pi_f) - \frac{1}{2}\sum_\text{$n$ odd}(\Pi_i + \Pi_f) G_{n\pi\overline{R}}(\Pi_i + \Pi_f)
\end{equation}
and the new momentum states are characterised by the original Neumann condition on the open string ends. To interpret this as strings moving in a single background we can introduce a Wilson line, $\Pi_i - e^{iA\int\!\ud X^0}\Pi_f$, with value $A=(\overline{R})^{-1}$.

In summary we have shown that the Schr\"odinger functional describing evolution through time t of second quantised strings can be written in terms of first quantised strings moving on the orbifold $\mathbb{S}^1/\mathbb{Z}_2$ and that the consequent T-duality interchanges t with 1/t and momentum fields with D$p$-brane backgrounds. BRST transformations describe reparametrisations of boundary data and for the open sting are sensitive to the Weyl anomaly even in the critical dimension. We have worked only in the free theory but interactions will be discussed in \cite{us}.

\end{document}